\documentclass[runningheads]{llncs}
\usepackage{graphicx}
\usepackage{amsmath}
\usepackage{amsfonts}
\usepackage{amssymb}
\usepackage{mathrsfs}
\usepackage{nicefrac}
\usepackage{multirow}

\newtheorem{defn}{Definition}

\begin{document}
\title{A Novel First-Order Autoregressive Moving Average Model to Analyze Discrete-Time Series Irregularly Observed.\thanks{Supported by CONICYT PFCHA/2015-21151457 and the ANID Millennium Science Initiative ICN12\_009, awarded to the Millennium Institute of Astrophysics.}}
\titlerunning{The iARMA model}
%
\author{C\'esar Ojeda\inst{1} \and Wilfredo Palma\inst{2} \and Susana Eyheramendy\inst{3} \and Felipe Elorrieta\inst{4}}
\authorrunning{C. Ojeda et al.}
%
\institute{Escuela de Estad\'istica, Universidad del Valle, Cali, Colombia\\ \email{cesar.ojeda@correounivalle.edu.co} \and
Millennium Institute of Astrophysics,  Santiago, Chile\\
\email{wilfredo.palma@gmail.com} \and
Faculty of Engineering and Sciences, Universidad Adolfo Iba\~nez, Santiago,  Chile\\ \email{susana.eyheramendy@uai.cl} \and Department of Mathematics and Computer Science, Universidad de Santiago, Chile\\ \email{felipe.elorrieta@usach.cl}}
\maketitle              
\begin{abstract}
A novel first-order autoregressive moving average model for analyzing discrete-time series observed at irregularly spaced times is introduced. 
Under Gaussianity, it is established that the model is strictly stationary and ergodic. In the general case, it is shown that the model is weakly stationary. The lowest dimension of the state-space representation is given along with the one-step linear predictors and their mean squared errors. The maximum likelihood estimation procedure is discussed, and their finite-sample behavior is assessed through Monte Carlo experiments. These experiments show that bias, root mean squared error, and coefficient of variation are smaller when the length of the series increases. Further, the method provides good estimations for the standard errors, even with relatively small sample sizes. Also, the irregularly spaced times seem to increase the estimation variability. The application of the proposed model is made through two real-life examples. The first is concerned with medical data, whereas the second describes an astronomical data set analysis.

\end{abstract}
\section{Introduction}

In statistics, time series analysis establishes a principal tool for studying time-ordered observations that are naturally dependent.
Nowadays, to study discrete-time series, many methods assume that time series are regularly observed; that is, the interval between observations is constant over time \cite{Brockwell1991,Hamilton1994,Box2016}. However, there are several fields as diverse as astronomy, climatology, economics, finance, medical sciences, geophysics, where time series are observed at irregularly spaced intervals \cite{moore1987experiences,Munoz1992,Belcher1994,kim2008fitting,Hand,CORDUAS20081860,Mudelsee2014,Babu,Eyheramendy2018,miller2019testing,Edelmann,refId0,ZHANG2020107019}. For example, \cite{Mudelsee2014} mentions that conventional time series analysis largely ignored irregularly spaced structures that climate time series has to consider.

The statistical analysis of irregular structures in time series poses several difficulties. First, the overwhelming majority of the available time series methods assume regularly observed data, as mentioned above. Second, when this assumption is dropped,  several technical problems arise including the issue of  formulating appropriate methodologies for carrying out  statistical inferences. Third, most of the currently available numerical algorithms for computing estimators and forecasts are based on the regularity of the data collection process. 

According to \cite{Jones1985}, irregularly spaced time series can occur in two different ways. On the one hand, data can be regularly spaced with missing observations. On the other hand, data can be truly irregularly spaced with no underlying sampling interval. Techniques considering discrete-time series in the presence of missing data have been studied, for instance \cite{Parzen1963,Jones1980,Dunsmuir1983,Reinsel1987}. Nevertheless, these techniques can not be applied if data are really irregularly spaced. When data are irregularly observed, it has been treated through two approaches. First, it could be transformed irregularly spaced time series into regularly spaced time series through interpolation to use traditional techniques. In \cite{Adorf1995}, can be found a summary of such transformations frequently used to analyze astronomical data. However, these interpolation methods typically produce bias (for instance, over smoothing), changing the dynamic of the process. Second, irregularly spaced time series can be treated as discrete realizations of a continuous stochastic process \cite{Robinson1977,Parzen1984,Thornton2013}. Nevertheless, continuous time series models tend to be computationally demanding and complicated (mostly due to the difficulty of estimating and evaluating them from discretely sampled data).
To analyze discrete-time series observed at irregularly spaced times directly, \cite{Eyheramendy2018} propose a first-order autoregressive model while \cite{ojeda2021irregularly} propose a first-order moving average model. Consequently, a novel model is proposed in this paper which allows for the treatment of moving averages and autoregressive structures with irregularly spaced discrete-times.

The remainder of the paper is organized as follows. Section \ref{sec:constructionModel} introduces the construction of the model. The model definition and its properties  it is given in Section \ref{sec:IARMA}. Also, this section provides the state-space representation of the model along with one-step linear predictors and their mean squared errors. The maximum likelihood estimation method is introduced in Section \ref{sec:IARMA-MLE}. The finite-samples behavior of this estimator is studied via Monte Carlo in Section \ref{sec:IARMA-MonteCarlo}. Two real-life data applications are discussed in Section \ref{sec:IMA-Application} while conclusions are given in Section \ref{sec:conclusions}.



\section{Model Formulation}\label{sec:constructionModel}

This section describes a stationary stochastic process with an autoregressive moving average structure that allows to consider irregularly spaced times. The pattern of irregular spacing is assumed to be independent of the stochastic process properties. Also, it is assumed that all joint moments up to order two are finite.

Let $\mathbb{T}={\{{t}_{n}\}}_{n\in{\mathbb{N}}^{+}}$ be a set of given times such that its consecutive differences, $\Delta_{n+1}=t_{n+1}-t_{n}$, are such that there is $\Delta_{L}>0$ such that $\Delta_{L}\leq\Delta_{n+1}$ for all $n$. Without loss of generality, it is assumed that  $\Delta_{L}=1$. Otherwise, each $t_{n}$ can be re-scaled by $\Delta_{L}$. These conditions are compatibles with any physical measurement and determine $\mathbb{T}$ as a discrete, and therefore countable subset of $\mathbb{R}$.

Let $\{\zeta_{t_{n}}\}_{t_{n}\in\mathbb{T}}$ be a sequence of uncorrelated--standardized random variables
and define the following sequence of real-valued random variables,
\[
X_{t_{1}}=\upsilon_{1}^{\nicefrac{1}{2}}\zeta_{t_{1}},\quad X_{t_{n+1}}=\phi^{\Delta_{n+1}}X_{t_{n}}+\upsilon_{n+1}^{\nicefrac{1}{2}}\zeta_{t_{n+1}}+\varpi_{n}\upsilon_{n}^{\nicefrac{1}{2}}\zeta_{t_{n}},
\]
where $0\leq\phi<1$; $\{\upsilon_{n}\}_{n\in{\mathbb{N}}^{+}}$ and $\{\varpi_{n}\}_{n\in{\mathbb{N}}^{+}}$ are time-varying sequences that characterize the moments of the process. Thus, for all $n$, $\textrm{E}(X_{t_{n}})=0$, $\textrm{Var}(X_{t_{1}})=\upsilon_{1}$, $\textrm{Var}(X_{t_{n+1}})=\phi^{2\Delta_{n+1}}\textrm{Var}(X_{t_{n}})+\upsilon_{n+1}+\varpi_{n}^{2}\upsilon_{n}+2\phi^{\Delta_{n+1}}\varpi_{n}\upsilon_{n}$, and
\begin{equation}
\textrm{Cov}(X_{t_{n}},X_{t_{n+k}})=\begin{cases}
\phi^{\Delta_{n+1}}\textrm{Var}(X_{t_{n}})+\varpi_{n}\upsilon_{n}, & k=1,\\
\phi^{\Delta_{n+k}}\textrm{Cov}(X_{t_{n}},X_{t_{n+k-1}}), & k\geq2.\label{eq:covarianceFunction}
\end{cases}
\end{equation}
By successive substitutions, for $k\geq2$, 
\[
\textrm{Cov}(X_{t_{n}},X_{t_{n+k}})=\phi^{t_{n+k}-t_{n+1}}\textrm{Cov}(X_{t_{n}},X_{t_{n+1}}).
\]

To obtain a stationary process, it is required that, for all $n$, $\textrm{Var}(X_{t_{n}})=\gamma_{0}$ and $\textrm{Cov}(X_{t_{n}},X_{t_{n+1}})=\gamma_{1,\Delta_{n+1}}$ with $\gamma_{0}$ time-independent and $\gamma_{1,\Delta_{n+1}}$ a function of $\Delta_{n+1}$. Thus,
\begin{gather}
\phi^{2\Delta_{n+1}}\gamma_{0}+\upsilon_{n+1}+\varpi_{n}^{2}\upsilon_{n}+2\phi^{\Delta_{n+1}}\varpi_{n}\upsilon_{n}=\upsilon_{1}=\gamma_{0},\;\textrm{and}\label{eq:Var-IARMA}\\
\phi^{\Delta_{n+1}}\gamma_{0}+\varpi_{n}\upsilon_{n}=\gamma_{1,\Delta_{n+1}}.\label{eq:Cov-IARMA}
\end{gather}
From \eqref{eq:Cov-IARMA},
\begin{equation}
\varpi_{n}=\frac{\gamma_{1,\Delta_{n+1}}-\phi^{\Delta_{n+1}}\gamma_{0}}{\upsilon_{n}}.\label{eq:IARMA-varpi}
\end{equation}
Replacing \eqref{eq:IARMA-varpi} into \eqref{eq:Var-IARMA},
\[
\upsilon_{n+1}=\gamma_{0}+\phi^{2\Delta_{n+1}}\gamma_{0}-2\phi^{\Delta_{n+1}}\gamma_{1,\Delta_{n+1}}-\frac{(\gamma_{1,\Delta_{n+1}}-\phi^{\Delta_{n+1}}\gamma_{0})^{2}}{\upsilon_{n}},\quad\textrm{with}\;\upsilon_{1}=\gamma_{0}.
\]
Also, since the process must be real-valued (i.e., without complex components), it is necessary that $\upsilon_{n}>0$, for all $n$. Thus, particular forms can be specified to $\gamma_{0}$ and $\gamma_{1,\Delta_{n+1}}$ that satisfy this condition to get the desired model. In this case, these forms are chosen to obtain the traditional ARMA(1,1) model when times are regularly observed. Consequently, consider $\gamma_{0}=\sigma^{2}\nicefrac{(1+2\phi\theta+\theta^{2})}{(1-\phi^{2})}$ and $\gamma_{1,\Delta_{n+1}}=\phi^{\Delta_{n+1}}\gamma_{0}+\sigma^{2}\theta^{\Delta_{n+1}}$ with $\sigma^{2}>0$ and $0\leq\phi,\theta<1$. Thence, $\varpi_{n}=\sigma^{2}\nicefrac{\theta^{\Delta_{n+1}}}{\upsilon_{n}}$ and
\[
\upsilon_{n+1}=\sigma^{2}\left(\frac{(1+2\theta\phi+\theta^{2})}{(1-\phi^{2})}(1-\phi^{2\Delta_{n+1}})-2\phi^{\Delta_{n+1}}\theta^{\Delta_{n+1}}-\frac{\sigma^{2}\theta^{2\Delta_{n+1}}}{\upsilon_{n}}\right)
\]
with $\upsilon_{1}=\gamma_{0}$.
To show that $\upsilon_{n}>0$, for all $n$, define
\[
c_{n+1}(\phi,\theta)=c_{1}(\phi,\theta)(1-\phi^{2\Delta_{n+1}})-2\phi^{\Delta_{n+1}}\theta^{\Delta_{n+1}}-\frac{\theta^{2\Delta_{n+1}}}{c_{n}(\phi,\theta)}
\]
with $c_{1}(\phi,\theta)=\nicefrac{(1+2\phi\theta+\theta^{2})}{(1-\phi^{2})}$. Hence, $\upsilon_{1}=\sigma^{2}c_{1}(\phi,\theta)$, $\upsilon_{n+1}=\sigma^{2}c_{n+1}(\phi,\theta)$,
and it would only be necessary to show that $c_{n}(\phi,\theta)>0$ for all $n$. Since $0\leq\phi,\theta<1$, then $c_{1}(\phi,\theta)\geq\nicefrac{1+\theta^{2}}{1-\phi^{2}}\geq1+\theta^{2}=c_{1}(\theta)>0$.
Also, since $1\leq\Delta_{n+1}$ for all $n$, then $\phi\theta\geq\phi^{\Delta_{n+1}}\theta^{\Delta_{n+1}}$ for all $n$. Thus,
\[
c_{n+1}(\phi,\theta)\geq1+\theta^{2}-\frac{\theta^{2\Delta_{n+1}}}{c_{n}(\phi,\theta)}=c_{n+1}(\theta).
\]
Here, $c_{n}(\phi,\theta)=c_{n}(\theta)$ since it is only a function of $\theta$. So, it suffices to show that $c_{n}(\theta)>0$ for all $n$ with $c_{1}(\theta)=1+\theta^{2}$ and $c_{n+1}(\theta)=c_{1}(\theta)-\nicefrac{\theta^{2\Delta_{n+1}}}{c_{n}(\theta)}$.
From \cite{Kilic2008}, the sequence $\{c_{n}(\theta)\}_{n\in{\mathbb{N}}^{+}}$ is known as a general backward continued fraction. In \cite{ojeda2021irregularly},
it is shown that assuming $1\leq\Delta_{n+1}$ for all $n$ and $0\leq\theta<1$, this sequence is strictly positive. Thus, $\upsilon_{n}>0$ for all $n$, and the desired model has been obtained.

\section{An Irregular Observed First-Order Autoregressive Moving Average Model}\label{sec:IARMA}

A novel stationary stochastic process with an autoregressive moving average structure allows considering irregularly observed times is defined. It is called irregularly observed first-order Autoregressive Moving Average (iARMA) model.

\begin{defn}[iARMA model]
\label{def:IARMA-Constructionist} Let $\{\varepsilon_{t_{n}}\}_{t_{n}\in\mathbb{T}}$ be a sequence of uncorrelated random variables with mean $0$ and variance $\sigma^{2}c_{n}(\phi,\theta)$ with $\sigma^{2}>0$, $0\leq\phi,\theta<1$, $c_{1}(\phi,\theta)=\frac{1+2\theta\phi+\theta^{2}}{1-\phi^{2}}$,
and
\[
c_{n+1}(\phi,\theta)=c_{1}(\phi,\theta)(1-\phi^{2\Delta_{n+1}})-2\phi^{\Delta_{n+1}}\theta^{\Delta_{n+1}}-\frac{\theta^{2\Delta_{n+1}}}{c_{n}(\phi,\theta)}.
\]
The process $\{X_{t_{n}}\}_{t_{n}\in\mathbb{T}}$ is said to be an iARMA process if $X_{t_{1}}=\varepsilon_{t_{1}}$, and
\begin{equation}
X_{t_{n+1}}=\phi^{\Delta_{n+1}}X_{t_{n}}+\varepsilon_{t_{n+1}}+\frac{\theta^{\Delta_{n+1}}}{c_{n}(\phi,\theta)}\varepsilon_{t_{n}}.\label{eq:IARMA}
\end{equation}
It is said that $\{X_{t_{n}}\}_{t_{n}\in\mathbb{T}}$ is an iARMA process with mean $\mu$ if $\{X_{t_{n}}-\mu\}_{t_{n}\in\mathbb{T}}$ is an iARMA process.
\end{defn}

In the iARMA model, when $\phi=0$, it is obtained the so-called iMA process \cite{ojeda2021irregularly}, while when $\theta=0$, the so-called iAR process \cite{Eyheramendy2018} is obtained. 

\subsection{Properties}

For the iARMA process, the mean and the autocovariance functions are
\begin{equation*}
\textrm{E}(X_{t_{n}})=0,\quad\textrm{and}\quad
\textrm{Cov}(X_{t_{n}},X_{t_{n+k}})=
\begin{cases}
	\sigma^{2}c_{1}(\phi,\theta), & k=0,\\
	\gamma_{1,\Delta_{n+1}}, & k=1,\\
	\phi^{t_{n+k}-t_{n+1}}\gamma_{1,\Delta_{n+1}}, & k\geq2,
\end{cases}
\end{equation*}
for all $n$, where $\gamma_{1,\Delta_{n+1}}=\sigma^{2}[\phi^{\Delta_{n+1}}c_{1}(\phi,\theta)+\theta^{\Delta_{n+1}}]$. The autocorrelation function is
\begin{equation*}
\textrm{Cor}(X_{t_{n}},X_{t_{n+k}})=
\begin{cases}
	1, & k=0,\\
	\rho_{1,\Delta_{n+1}}, & k=1,\\
	\phi^{t_{n+k}-t_{n+1}}\rho_{1,\Delta_{n+1}}, & k\geq2,
\end{cases}
\end{equation*}
for all $n$, where $\rho_{1,\Delta_{n+1}}=\phi^{\Delta_{n+1}}+\nicefrac{\theta^{\Delta_{n+1}}}{c_{1}(\phi,\theta)}$. Since the process has a constant mean and a covariance function that depends only on the time differences, the process is weakly stationary. In particular, if $\{\varepsilon_{t_{n}}\}_{t_{n}\in\mathbb{T}}$ are independent random variables each $\textrm{N}(0,\sigma^{2}c_{n}(\phi,\theta))$, then the iARMA process would be a weakly stationary Gaussian process, and therefore strictly stationary.
Also, when $\Delta_{n+1}=1$ for all $n$, it is obtained the traditional ARMA(1,1) process.

Now, from \eqref{eq:IARMA}, consider $Y_{t_{n+1}}=\varepsilon_{t_{n+1}}+[\nicefrac{\theta^{\Delta_{n+1}}}{c_{n}(\phi,\theta)}]\varepsilon_{t_{n}}$ with $\textrm{Var}(Y_{t_{n+1}})=\sigma^{2}[c_{1}(\phi,\theta)(1-\phi^{2\Delta_{n+1}})-2\phi^{\Delta_{n+1}}\theta^{\Delta_{n+1}}]$. Hence, $X_{t_{n+1}}=\phi^{\Delta_{n+1}}X_{t_{n}}+Y_{t_{n+1}}$ for all $n$,
with $X_{t_{1}}=\varepsilon_{t_{1}}$ and $\textrm{Cov}(X_{t_{n}},Y_{t_{n+1}})=\sigma^{2}\theta^{\Delta_{n+1}}$. By successive substitutions,
\begin{equation}
X_{t_{n+1}}=\phi^{t_{n+1}-t_{1}}\varepsilon_{t_{1}}+\sum_{j=1}^{n}\phi^{t_{n+1}-t_{j+1}}Y_{t_{j+1}}.\label{eq:iARMA-Wold-representation}
\end{equation}
Consequently, for larger $n$, the initial condition effect vanishes. Thus, the process \textquotedblleft forget\textquotedblright{} its initial starting value. Also, from \eqref{eq:iARMA-Wold-representation}, $X_{t_{n}}$ can be expressed as a function of $\{\varepsilon_{t_{j}}\}_{j=1}^{n}$, for each $n$. Then, under independence between these errors, $X_{t_{n}}$ is ergodic \cite{Stout1974}.

\subsection{State-Space Representation}\label{sec:IARMA-StateSpace}

From Definition \ref{def:IARMA-Constructionist}, it is presented a state-space representation of the model \eqref{eq:IARMA}. It enables the application of the Kalman filter for prediction and allows the maximum likelihood estimation, see \cite{Harvey1989}. This representation has the lowest dimension of the state vector and is given by
\[
X_{t_{n}}=\alpha_{t_{n}}+\varepsilon_{t_{n}},\quad\alpha_{t_{1}}=0,\;\alpha_{t_{n+1}}=\phi^{\Delta_{n+1}}\alpha_{t_{n}}+\left(\phi^{\Delta_{n+1}}+\frac{\theta^{\Delta_{n+1}}}{c_{n}(\phi,\theta)}\right)\varepsilon_{t_{n}}.
\]
In this representation, measurement and transition equation disturbances are correlated. From \cite{Harvey1989}, these equations can be transformed into a new system with disturbances uncorrelated, which are
\begin{equation}
X_{t_{n}}=\alpha_{t_{n}}+\varepsilon_{t_{n}},\quad\alpha_{t_{1}}=0,\;\alpha_{t_{n+1}}=\left(\phi^{\Delta_{n+1}}+\frac{\theta^{\Delta_{n+1}}}{c_{n}(\phi,\theta)}\right)X_{t_{n}}-\frac{\theta^{\Delta_{n+1}}}{c_{n}(\phi,\theta)}\alpha_{t_{n}}. \label{eq:IARMA-TranEqUncorrelated}
\end{equation}
The inclusion of $X_{t_{n}}$ in \eqref{eq:IARMA-TranEqUncorrelated} does not affect the Kalman filter, as $X_{t_{n}}$ is known at time $t_{n}$.

\subsection{Prediction}\label{sec:IARMA-Prediction}

Using the innovations algorithm \cite{Brockwell1991}, the one-step linear predictors for the iARMA model are $\hat{X}_{t_{1}}(\phi,\theta)=0$, with mean squared error $\textrm{E}\{(X_{t_{1}}-\hat{X}_{t_{1}}(\phi,\theta))^{2}\}=\sigma^{2}c_{1}(\phi,\theta)$, and
\[
\hat{X}_{t_{n+1}}(\phi,\theta)=\phi^{\Delta_{n+1}}X_{t_{n}}+\frac{\theta^{\Delta_{n+1}}}{c_{n}(\phi,\theta)}(X_{t_{n}}-\hat{X}_{t_{n}}(\phi,\theta)),\;n\geq1,
\]
with mean squared errors $\textrm{E}\{(X_{t_{n+1}}-\hat{X}_{t_{n+1}}(\phi,\theta))^{2}\}=\sigma^{2}c_{n+1}(\phi,\theta)$.


\section{Maximum Likelihood Estimation}\label{sec:IARMA-MLE}

Let $X_{t}$ be observed at points $t_{1},\ldots,t_{\textrm{N}}$. The log-likelihood under Gaussianity is
\[
-\frac{\textrm{N}}{2}\ln{2\pi}-\frac{\textrm{N}}{2}\ln{\sigma^{2}}-\frac{1}{2}\sum_{n=1}^{\textrm{N}}\ln{c_{n}(\phi,\theta)}-\frac{1}{2}\sum_{n=1}^{\textrm{N}}\frac{(X_{t_{n}}-\hat{X}_{t_{n}}(\phi,\theta))^{2}}{\sigma^{2}c_{n}(\phi,\theta)},
\]
where $\phi,\theta$ and $\sigma^{2}$ are any admissible parameter values. Now, optimizing it for $\sigma^{2}$, replacing the optimum into the log-likelihood, and organizing terms, it is obtained the reduced likelihood $q_{\textrm{N}}(\phi,\theta)=\ln{\hat{\sigma}_{\textrm{N}}^{2}(\phi,\theta)}+\nicefrac{1}{\textrm{N}}\sum_{n=1}^{\textrm{N}}\ln{ c_{n}(\phi,\theta)}$
with $\hat{\sigma}_{\textrm{N}}^{2}(\phi,\theta)=\nicefrac{1}{\textrm{N}}\sum_{n=1}^{\textrm{N}}\nicefrac{(X_{t_{n}}-\hat{X}_{t_{n}}(\phi,\theta))^{2}}{c_{n}\left(\phi,\theta\right)}$. 
The maximum likelihood estimates of $\phi$ and $\theta$, denoted as $\hat{\phi}_{\textrm{N}}$ and $\hat{\theta}_{\textrm{N}}$, respectively, are the values minimizing $q_{\textrm{N}}(\phi,\theta)$. The estimate of $\sigma^{2}$ is $\hat{\sigma}_{\textrm{N}}^{2}=\sigma_{\textrm{N}}^{2}(\hat{\phi}_{\textrm{N}},\hat{\theta}_{\textrm{N}})$. The optimization can be done through the method proposed by \cite{Byrd1995}, which allows general box constraints. Specifically, $q_{\textrm{N}}(\phi,\theta)$ can be minimized under the constraint $0\leq\phi,\theta<1$. Also, this method allows for finding the numerically differentiated Hessian matrix at the solution given. Solving it, and according to \cite{Hamilton1994}, estimated standard errors can be obtained.

\section{Monte Carlo Experiments}\label{sec:IARMA-MonteCarlo}

This section provides a Monte Carlo study that assesses the finite-sample performance of the Maximum Likelihood (ML) estimator. The simulation consider $\sigma^{2}=1$, $\phi\in\left\{ 0.5\right\} $, $\theta\in\left\{ 0.1,0.5,0.9\right\}$, and $\textrm{N}\in\left\{ 100,500,1500\right\}$, where $\textrm{N}$ represents the length of the series. Furthermore, $\textrm{M}=1000$ trajectories are simulated, and for each, $\phi$ and $\theta$ are estimated. It is regarded as regular ($\Delta_{n}=1$ for $n=2,\ldots,\textrm{N}$) as well as irregular spaced times, where $\Delta_{n}\overset{\textrm{ind}}{\sim}1+\textrm{exp}(\lambda=1)$, for $n=2,\ldots,\textrm{N}$.
Now, let $\hat{\phi}_{m}$ and $\hat{\theta}_{m}$ be the ML estimations for the $m$-th trajectory with $\widehat{\textrm{se}}(\hat{\phi}_{m})$ and $\widehat{\textrm{se}}(\hat{\theta}_{m})$ their estimated standard errors. These standard errors are estimated through the curvature of the likelihood surface at $\hat{\phi}_{m}$ and $\hat{\theta}_{m}$ (see, Section \ref{sec:IARMA-MLE}). As a summary of these quantities, the mean value of the $\textrm{M}$ maximum likelihood estimations are computed. For example, for the moving average parameter, $\hat{\theta}=\nicefrac{1}{\textrm{M}}\sum_{m=1}^{\textrm{M}}\hat{\theta}_{m}$ and $\widehat{\textrm{se}}(\hat{\theta})=\nicefrac{1}{\textrm{M}}\sum_{m=1}^{\textrm{M}}\widehat{\textrm{se}}(\hat{\theta}_{m})$.

\subsection{Performance Measures}

As a measure of estimator performance, Root Mean Square Error (RMSE) and Coefficient of Variation (CV) are considered. For example, for the ML estimator for $\theta$, $\textrm{RMSE}_{\hat{\theta}}=(\widehat{\textrm{se}}(\hat{\theta})^{2}+\textrm{bias}_{\hat{\theta}}^{2})^{\nicefrac{1}{2}}$, and $\textrm{CV}_{\hat{\theta}}=\nicefrac{\widehat{\textrm{se}}(\hat{\theta})}{|\hat{\theta}|}$,
where $\textrm{bias}_{\hat{\theta}}=\hat{\theta}-\theta$. Furthermore, as an approximate variance of the estimator, $\widetilde{\textrm{se}}^{2}(\hat{\theta})=\nicefrac{1}{\textrm{M}-1}\sum_{m=1}^{\textrm{M}}(\hat{\theta}_{m}-\hat{\theta})^{2}$ is used.
Finally, according to \cite{Koehler2009}, the Monte Carlo Error (MCE) is estimated  for every simulation via asymptotic theory through $\nicefrac{\widetilde{\textrm{se}}(\hat{\theta})}{\sqrt{M}}$.
Remember that the MCE is a estimation of the standard deviation of the Monte Carlo estimator, taken across repetitions of the simulation, where each simulation is based on the same design and consists of $M$ replications.


\subsection{Simulation Results}

Table \ref{IARMA-ML-MC-Table-IrregularSpaced} shows the performance measures of the estimator for maximum likelihood method.
Bias, RMSE and CV are smaller when $\textrm{N}$ increases as expected. Also, the method provides good estimations for the standard error, even with relatively small sample sizes. Furthermore, although it is not shown,  comparing these results with the one obtained assuming regularly spaced times (the conventional first-order ARMA model), the irregularly spaced times seem to increase the estimation variability. 
\begin{center}
\begin{table}
\begin{centering}
\begin{tabular}{|c|c|c|c|c|c|c|c|}
\hline 
N & $\theta$ & $\hat{\theta}$ & $\widehat{\textrm{se}}(\hat{\theta})$ & $\widetilde{\textrm{se}}(\hat{\theta})$ & $\textrm{bias}_{\hat{\theta}}$ & $\textrm{RMSE}_{\hat{\theta}}$ & $\textrm{CV}_{\hat{\theta}}$\tabularnewline
\hline 
\multirow{3}{*}{$100$} & $0.1$ & $0.294$ & $0.245$ & $0.245$ & $0.194$ & $0.312$ & $0.835$\tabularnewline
 & $0.5$ & $0.500$ & $0.252$ & $0.263$ & $0.000$ & $0.252$ & $0.505$\tabularnewline
 & $0.9$ & $0.796$ & $0.232$ & $0.228$ & $-0.104$ & $0.255$ & $0.292$\tabularnewline
\hline 
\multirow{3}{*}{$500$} & $0.1$ & $0.192$ & $0.158$ & $0.179$ & $0.092$ & $0.183$ & $0.827$\tabularnewline
 & $0.5$ & $0.501$ & $0.149$ & $0.160$ & $0.001$ & $0.149$ & $0.298$\tabularnewline
 & $0.9$ & $0.885$ & $0.090$ & $0.094$ & $-0.015$ & $0.091$ & $0.102$\tabularnewline
\hline 
\multirow{3}{*}{$1500$} & $0.1$ & $0.131$ & $0.102$ & $0.116$ & $0.031$ & $0.106$ & $0.780$\tabularnewline
 & $0.5$ & $0.499$ & $0.094$ & $0.098$ & $-0.001$ & $0.094$ & $0.188$\tabularnewline
 & $0.9$ & $0.895$ & $0.050$ & $0.049$ & $-0.005$ & $0.050$ & $0.056$\tabularnewline
\hline 
N & $\phi$ & $\hat{\phi}$ & $\widehat{\textrm{se}}(\hat{\phi})$ & $\widetilde{\textrm{se}}(\hat{\phi})$ & $\textrm{bias}_{\hat{\phi}}$ & $\textrm{RMSE}_{\hat{\phi}}$ & $\textrm{CV}_{\hat{\phi}}$\tabularnewline
\hline 
$100$ & \multirow{3}{*}{$0.5$} & $0.448$ & $0.155$ & $0.167$ & $-0.052$ & $0.163$ & $0.346$\tabularnewline
$500$ &  & $0.488$ & $0.076$ & $0.079$ & $-0.012$ & $0.077$ & $0.156$\tabularnewline
$1500$ &  & $0.497$ & $0.046$ & $0.048$ & $-0.003$ & $0.046$ & $0.092$\tabularnewline
\hline
\end{tabular}
\par\end{centering}
\caption{Monte Carlo results for the irregularly spaced time case. The maximum MCE
estimated (in all simulations) is $0.008$. When $\phi=0.5$, we use $\theta=0.5$.}
\label{IARMA-ML-MC-Table-IrregularSpaced}
\end{table}
\par\end{center}

\section{Applications}\label{sec:IMA-Application}

This section illustrates the application of the proposed time series model to two real-life datasets. The first example is concerned with medical data whereas the second application describes the analysis of an astronomical data set.

\subsection{Lung Function of an Asthma Patient}

\begin{figure}[h]
\centering
\includegraphics[scale=0.65]{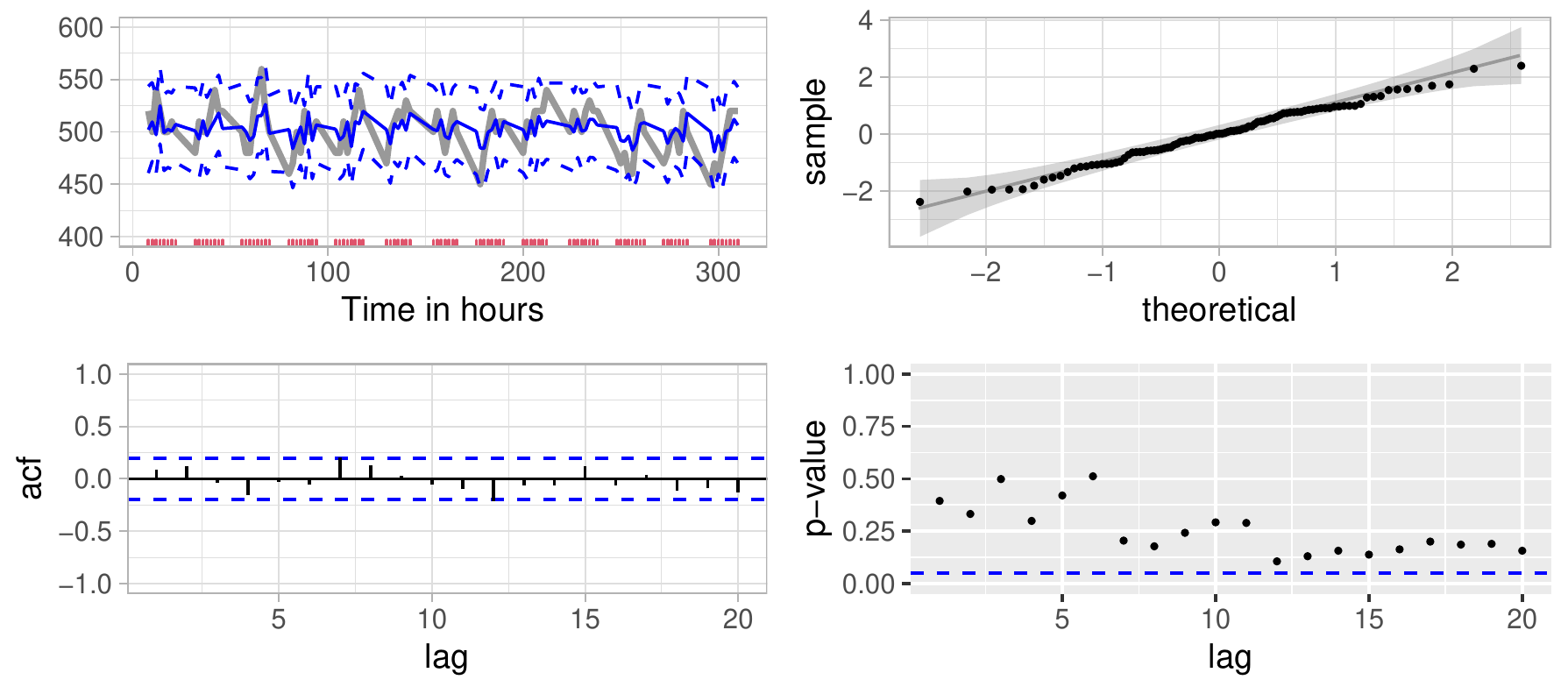}
\caption{On the left-top, the lung function of an asthma patient with the predicted values and their respective variability bands. For the standardized residuals: on the right-top, the quantile-quantile plot with normality reference bands \cite{Nair1982}; on the bottom-left, the autocorrelation function estimated; on the bottom-right, the Ljung-Box test for randomness.}
\label{medical-application}
\end{figure}

In \cite{Belcher1994}, it is analyzed measurements of the lung function of an asthma patient. The  observations are collected mostly at 2 hour time intervals but with irregular gaps (see the unequal spaced of tick marks in Figure \ref{medical-application}). However, as it was shown in \cite{Wang2013}, the trend component (obtained by decomposing original time series into trend, seasonal, and irregular components via the Kalman smoother) exhibits structural changes after 100$th$ observation. Thus,  the first 100 observations are considered here to analyze such a phenomenon. Below,  the ML estimates are reported along with  their respective estimated standard errors. Here, the autoregressive estimate is not significant (not shown), but the others estimates are significant at the 5\% significance level suggesting an iMA model.
\[
\hat{\theta}=0.853\quad\widehat{\textrm{se}}(\hat{\theta})=0.069\quad\hat{\sigma}^{2}=258.286\quad\widehat{\textrm{se}}(\hat{\sigma}^{2})=36.537
\]

From Figure \ref{medical-application}, the fit seems adequate. Also, the standardized residuals seem to follow a standard normal distribution. Furthermore, this figure shows the ACF estimated and the results from a Ljung-Box test for the standardized residuals. Observe that the residuals satisfy the white noise test at the $5$\% significance level. Note that, since the standardized residuals are assumed  to be realizations of a random sample, its correlation structure does not depend on the irregularly spaced between observations. Thus, unlike the original time series, the  ACF and the Ljung-Box test can be applied to the standardized residuals.


\subsection{Light Curve of an Astronomical Object}\label{sec:IARMA-Application}

\begin{figure}[h]
\centering
\includegraphics[scale=0.6]{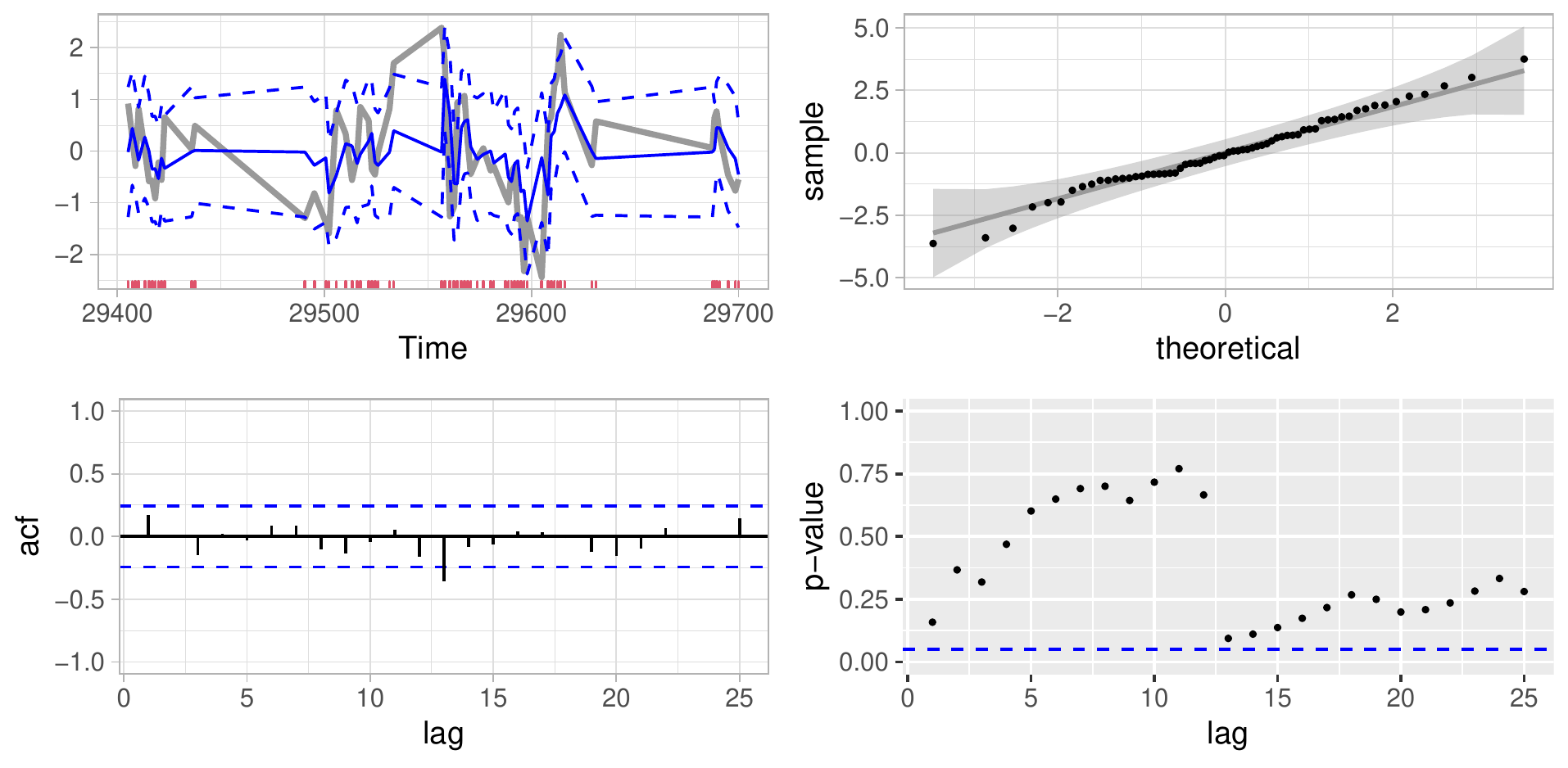}
\caption{On the left-top, the light curve of the Blazar object with the predicted values and their respective variability bands. For the standardized residuals: on the right-top, the quantile-quantile plot with normality reference bands \cite{Nair1982}; on the bottom-left, the autocorrelation function estimated; on the bottom-right, the Ljung-Box test for randomness.}
\label{astronomical-application}
\end{figure}

In astronomy the study of the temporal behavior of the brightness of different objects is a matter of interest. The time series of the brightness of an astronomical object is called as light curve. Light curves are commonly measured at irregular times. In this work, it is also assess the performance of the iARMA model in a light curve of an astronomical object. The light curve that it is used was observed with the Zwicky Transient Facility (ZTF), see \cite{Bellm2018}, and belongs to a Blazar astronomical object coded as ``ZTF18aabxyhf''. The time series data of this Blazar were processed by the ALeRCE broker \cite{ALeRCE}. The light curve of this object has 65 measurements of the brightness of this object in a range of approximately 584 days. The average gap of the observations of this light curve is 9.13 days. The iARMA model parameters were estimated via maximum likelihood method in this light curve yielding the following results:
\begin{gather*}
\hat{\phi}=0.702\quad\widehat{\textrm{se}}(\hat{\phi})=0.112\quad\hat{\theta}=0.682\quad\widehat{\textrm{se}}(\hat{\theta})=0.366\\
\hat{\sigma}^{2}=0.209\quad\widehat{\textrm{se}}(\hat{\sigma}^{2})=0.064.
\end{gather*}
According to this results, both the $\phi$ and $\theta$ parameters are significative at 10\% level.  Furthermore, in Figure \ref{astronomical-application} it is shown that the residuals of the iARMA model do not hold an autocorrelation structure. In other words, the iARMA explains all the time dependence of the observed light curve.  Also, the standardized residuals seem to follow a standard normal distribution.

\section{Conclusions}\label{sec:conclusions}

An irregularly observed first-order autoregressive moving average model was proposed that allows treating first-order autoregressive moving averages structures with irregularly spaced times. It is established that, under Gaussianity, the model is strictly stationary and ergodic. The lowest dimension of the state-space representation along with the one-step linear predictors and its mean squared errors were given. Through of a Monte Carlo study, for the ML estimation method, it is shown that bias, RMSE and CV are smaller when $\textrm{N}$ increases. Also, the method provides good estimations for the standard errors, even with relatively small sample sizes. Furthermore, the irregularly spaced times seem to increase the estimation variability. It should be noted that, despite not being presented here,  the same Monte Carlo study was done for a proposed bootstrap estimation method. It showed a consistent behavior similar to what was found for the ML method. Finally, the practical application of the proposed methodology is illustrated by means of two real-life data examples involving medical and astronomical time series.


%
%
%
\bibliographystyle{splncs04}
\bibliography{samplepaper}

\end{document}